\documentclass[]{interact}

\allowdisplaybreaks

\usepackage{epstopdf}
\usepackage[caption=false]{subfig}

\usepackage{amssymb}
\usepackage{amsmath}
\usepackage[monochrome]{xcolor}
\usepackage{graphicx}
\usepackage{natbib}
\usepackage{epstopdf}
\usepackage{multicol}
\usepackage{float}
\usepackage{cite}
\usepackage{url}
\usepackage[boxed]{algorithm}
\usepackage{algorithmic}

\usepackage{balance}

\makeatletter
\newcommand{\figcaption}{\def\@captype{figure}\caption}
\makeatother

\theoremstyle{plain}
\newtheorem{theorem}{Theorem}[section]
\newtheorem{lemma}[theorem]{Lemma}
\newtheorem{corollary}[theorem]{Corollary}

\newtheorem{assumption}[theorem]{Assumption}

\theoremstyle{definition}

\theoremstyle{remark}
\newtheorem{remark}{Remark}

\begin{document}


\title{Observer-Based Distributed Leader-Follower Tracking Control: A New Perspective and Results}

\author{
\name{Chuan Yan\textsuperscript{a}\thanks{CONTACT Huazhen Fang. Email: fang@ku.edu} and Huazhen Fang\textsuperscript{a}}
\affil{\textsuperscript{a}Department of Mechanical Engineering, University of Kansas, Lawrence, KS, USA}}

\maketitle

\begin{abstract}
Leader-follower tracking control design has received significant attention in recent years due to its important and wide applications.
Considering a multi-agent system composed of a leader and multiple followers, this paper proposes and investigates a new perspective into this problem: can we enable a follower to estimate the leader's driving input and leverage it to develop new observer-based tracking control approaches? With this motivation, we develop an input-observer-based leader-follower tracking control framework, which features distributed input observers that allow a follower to locally estimate the leader's input toward enhancing tracking control. This work first studies the first-order tracking problem. It then extends to the more sophisticated case of second-order tracking and considers a challenging situation when the leader's and followers' velocities are not measured. The proposed approaches exhibit interesting and useful advantages as revealed by a comparison with the literature. Convergence properties of the proposed approaches are rigorously analyzed. Simulation results further illustrate the efficacy of the proposed perspective, framework and approaches.

\end{abstract}


\section{Introduction}


A multi-agent system (MAS) is a system composed of multiple agents interacting with each other, which allows for inter-agent connection and operation, distributed computation and control, and collective response to environment or external conditions~\citep{ferber:1999:AWR}. With a wide application spectrum in scientific, commercial and military sectors, it has attracted considerable attention and research from different communities. Coordinated control design is central to the successful accomplishment of MAS tasks, which thus has emerged as an active research field in the systems and control community. This field includes a broad range of problems of interest, including group consensus, synchronization, rendezvous, coverage control and leader-follower tracking, see~\citep{wang:2010:TAC,Wu:SCL:2011,Li:AUTO:2015,
yuwang:2010:SCL,yu:2016:AJC,Yang:AUTO:2014,
mastellone:2008:IJRR,lin:2005:TAC,Mou:TAC:2016,Dorfler:PNAS:2013,jadbabaie:2004:ACC,lin:2007:SIAM,
Cortes:TRA:2004,schwager:2009:IJRR,yoo:2013:AUTO,li:2013:IJRNC,hu:2014:aut}. Among them, leader-follower tracking often plays a critical role in missions ranging from rescue and search to delivery, surveillance, reconnaissance and mapping~\citep{Lewis:Springer:2014}.


In a leader-follower MAS, a swarm of agents referred to as followers interchange information and apply local control to cooperatively track a leader agent's behavior. The past decade has witnessed a growing amount of research on  control design to accomplish this objective, e.g.,~\citep{hong:2008:distributed,li:2011:AUT,Zhang:IJC:2013,cao:2015:SCL,zhu:2010:AUT,
hu:2010:AUT,hu:2015:CNSNS,Yan:ACC:2018,Yan:CEP:2018} and the references therein. Like other MAS control problems, this problem faces a fundamental challenge that a follower has limited access to information about the other agents (leader and other followers). A primary reason is that information exchange across  an MAS is distributed and localized. That is, a follower can only exchange information with its neighbors, and only a subset of the followers can directly communicate with the leader. Adding to this situation, an agent may be unable to measure all of its state variables because sensing devices can be unavailable or too expensive. Consequently, significant research effort has been devoted to observer-based control design, in which followers run observers to estimate the leader's and/or their own state for the purpose of control. The literature includes two main types of approaches in this regard. {\color{blue} The first type is  about velocity or position observers designed for  MASs based on a first- or second-order  model, and the second type about state observers for MASs characterized by state-space models.}
\begin{itemize}
\item {\color{blue} {\em Velocity/position-observer-based control}. For a second-order MAS, the leader's velocity is useful for tracking control but inaccessible to followers when agents do not have velocity sensors. A lead is taken in~\citep{hong:2008:distributed} with the development of a distributed observer that allows a follower to estimate the leader's velocity. The notion   is extended in~\citep{Chen:IJC:2014} to achieve tracking control in a sampled-data setting and in~\citep{li:2011:AUT} to enable finite-time leader-follower consensus. In~\citep{hu:2015:CNSNS}, an observer is proposed for a follower to estimate its relative velocity with respect to the leader. Observer design can also be leveraged to estimate followers' velocties. In~\citep{Xu:Neurocomputing:2015}, a local velocity observer is proposed so that a follower can reconstruct its own velocity. A similar problem is investigated in~\citep{Zhang:SCL:2014}. The approach therein includes an observer, which, though not making explicit velocity estimation, is still meant to make up for the absent velocity information.
Position-observer-based tracking control for a first-order MAS is studied in~\citep{Wang:CTT:2016}, in which a position observer is designed to allow followers to estimate the leader's position. However, it requires the leader's control law to take a specific linear form and be known by all the followers to ensure effective position estimation and tracking.
}

\item {\em State-observer-based control.} When agents have dynamics modeled in the linear state-space form, a state observer is often needed to achieve output-feedback control. A Luenberger-like observer in~\citep{Zhang:TAC:2011} is proposed for a follower to estimate its local state, which adopts state correction using the follower's  output estimation error relative to its neighbors'. Akin to this, state observers are designed and used in~\citep{Xu:Neurocomputing:2013} for  tracking control in the presence of switching topology and in~\citep{Shi:IJRNC:2017,Peng:TNNLS:2014} for leader-follower synchronization with uncertainties.

\end{itemize}

The studies surveyed above not only provide a wealth of results  regarding observer-based tracking control but also show the significance and potential  of observers for this control problem. It is noted, however, that the observer design has been almost solely focused on estimating the state variables (e.g., velocity of a second-order agent or  state vector of a state-space agent), either the leader's or a follower's. By comparison, estimation of the leader's input has received far less attention, even though it is evident that knowledge of a leader's maneuver input, if available in real time, can critically help a follower keep tracking the leader. Hence, we consider a new {\em perspective} to investigate leader-follower tracking control by developing distributed input observers that can enable every follower to estimate the leader's input. Since the  input observers can bring a follower an awareness of the leader's maneuvers, the tracking control can be hopefully enhanced.

This perspective leads us to make a two-fold contribution through  this work. First, we propose a novel input-observer-based tracking control framework. As a distinguishing feature, this framework includes  distributed input observers run by followers to  estimate the leader's control input. Compared to~\citep{Wang:CTT:2016}, such observers would neither require the leader's control law to take a special form nor demand it to be known by every follower. Second, following this framework, we systematically develop new tracking control approaches for both first- and second-order MASs. This involves the development of distributed input observers, together with some other observers for position or velocity estimation, and integrates them into tracking control laws. Theoretical analysis proves the effectiveness of the proposed approaches, which is further validated by simulation results. The proposed approaches will bring important benefits for tracking control, e.g., loosening some long-held assumptions and reducing the need for sensing devices, with a detailed discussion offered in the later sections.

The rest of this paper is organized as follows. Section~\ref{notation} summarizes the notation used in this paper. Section~\ref{first-order} formulates the problem of interest and presents the input-observer-based framework design for first-order leader-follower tracking. 
Section~\ref{second-order} studies the input-observer-based tracking for the second-order case. Simulation studies are offered in Section~\ref{simulation} to illustrate the proposed approaches. Finally, Section~\ref{conclusion} gathers our
concluding remarks.

\section{Notation}\label{notation}
The notation throughout this paper is standard. The set of real numbers is denoted by $\mathbb{R}$. The one norm of a vector is denoted as $\|\cdot\|_1$. We let $\det(\cdot)$ represent the determinant of a matrix and $\mathbf 1$ denote a column vector with all elements equal to 1. Matrices,
if their dimensions are not indicated explicitly, are assumed to be
compatible in algebraic operations. We use a graph to describe the topological structure for information exchange among the leader and followers. First, consider a network composed of $N$ independent followers. The interaction topology is modeled as an undirected graph. The follower graph is expressed as $\mathcal{G}=(\mathcal{V},\mathcal{E})$, where $\mathcal{V}=\{1, 2, \cdots, N\}$ is the node set and the edge set $\mathcal{E}\subseteq \mathcal{V}\times \mathcal{V}$ contains unordered pairs of nodes. A path is a sequence of connected edges in a graph.
The neighbor set of agent $i$ is denoted as $\mathcal{N}_i$, which includes all the agents in communication with it.
The adjacency matrix of $\mathcal{G}$ is $A=[a_{ij}]\in \mathbb{R}^{N \times N}$, which has non-negative elements. The element $a_{ij}>0 $ if and only if $(i,j)\in\mathcal{E}$, and moreover, $a_{ii}=0$ for all $i \in \mathcal{V}$. For the Laplacian matrix
$L=[l_{ij}]\in\mathbb{R}^{N \times N}$, $l_{ij}=-a_{ij}$ if $i\neq j$ and $ l_{ii}=\sum_{k\in \mathcal{N}_{i}}a_{ik}$.
The leader is numbered as vertex $0$, and information can be exchanged between the leader and its neighbors. Then, we have a graph $\bar{\mathcal{G}}$ , which consists of graph $\mathcal{G}$, vertex $0$ and edges from the vertex $0$ (i.e., the leader) to its neighbors. The leader is globally reachable in $\bar{\mathcal{G}}$ if there is a path from node 0 to every node $i$ in $\mathcal{G}$. In order to express the graph $\bar{\mathcal{G}}$ more precisely, we denote the leader adjacency matrix associated with $\bar{\mathcal{G}}$ by $B=\mathrm{diag}(b_1,\ldots,b_N)$, where $b_i>0$ if the leader is a neighbor of agent $i$ and $b_i=0$ otherwise. The following lemmas will be useful.

\begin{lemma}\citep{ren:2010:SSBM}\label{L-notation}
The Laplacian matrix $L(\mathcal{G})$ has at least one zero eigenvalue, and all the nonzero eigenvalues are
positive. Furthermore, $L(\mathcal{G})$ has a
simple zero eigenvalue and all the nonzero eigenvalues are positive if and only if $\mathcal{G}$ is connected.
\end{lemma}

\begin{lemma}\citep{hu:2007:PASMA}\label{H-notation}
The matrix $H=lB+L$ is positive stable {\color{blue}(i.e., all the eigenvalues have a positive real part)}, where $l>0$ is a positive coefficient,
if and only if vertex 0 is globally reachable.
\end{lemma}

\section{First-order Leader-follower Tracking}\label{first-order}
In this section, we first formulate the problem of first-order leader-follower tracking to be considered. Then, we develop an input-observer-based tracking control approach with convergence proof provided. In the end, the results are extended to a simplified yet meaningful case.

\subsection{Problem Formulation and Proposed Algorithm}
Consider a leader-follower MAS, where the followers are expected to track the leader's trajectory to accomplish an assigned mission. During the tracking process, the leader and followers maintain communication according to a pre-specified network topology to exchange their state information. Leveraging the information received, {\color{blue}the followers can determine their control inputs and then steer themselves to track the leader}. Suppose that the leader is numbered as $0$ and that the $N$ followers are numbered from $1$ to $N$.
Their dynamics is given by
\begin{align}\label{follower-dynamics-first}
\dot{x}_{i}=u_{i},\quad x_{i}\in \mathbb{R}, \quad i=0,1,\ldots,N,
\end{align}
where $x_{i}$ is the position and $u_{i}$ the control input.
Given this problem setting, the aim is to design $u_i$ for $i=1,2,\ldots,N$ such that follower $i$ can asymptotically track the leader, i.e., $\lim_{t\rightarrow \infty}|x_{i}(t)-x_{0}(t)|=0$.

\begin{figure}[t]
\begin{center}
\includegraphics[scale=0.4]{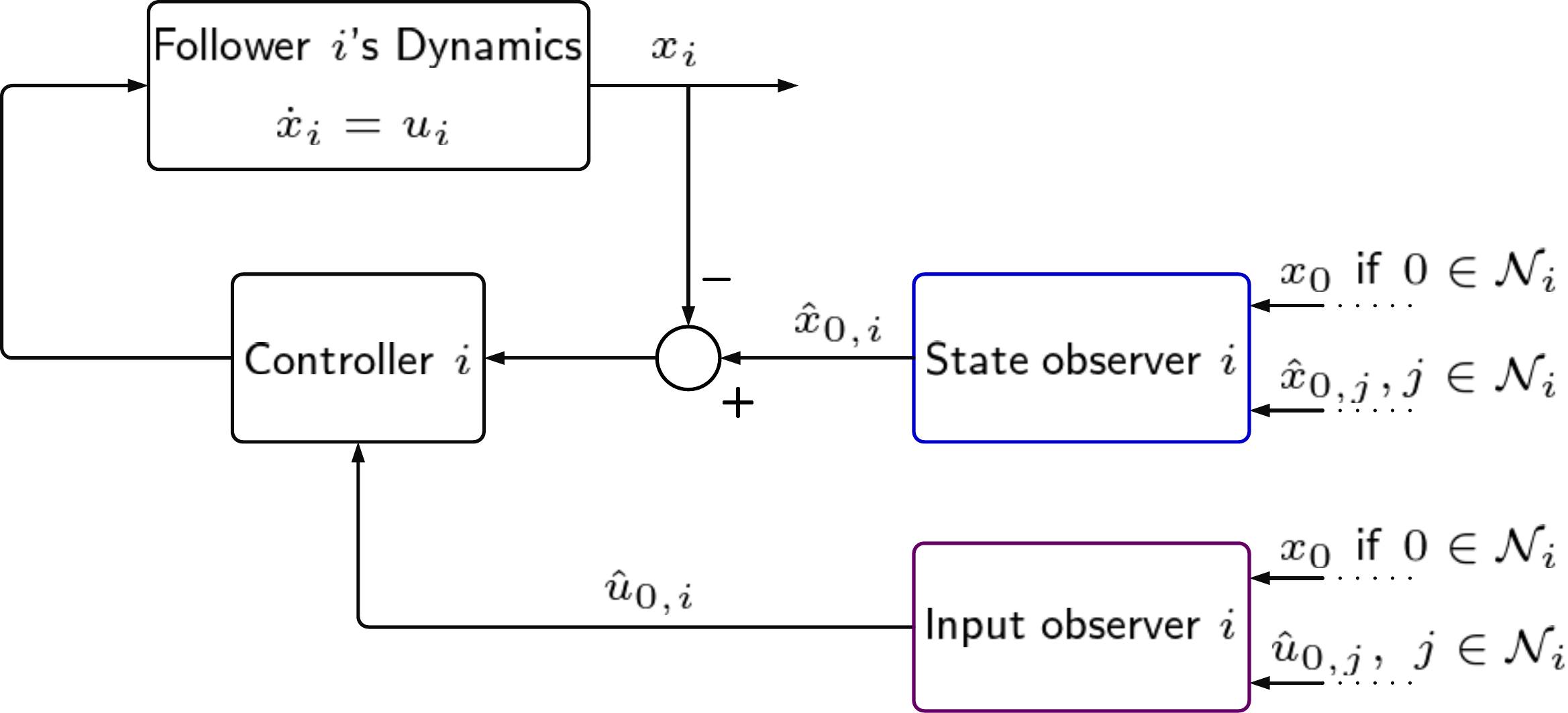}
 \caption{Input-observer-based framework for leader-follower tracking.
}
\label{diagram-V1}
\end{center}
\end{figure}

To achieve the above aim, we develop an input-observer-based tracking control design methodology. As a first step, we propose the conceptual design of a linear continuous controller. Because the leader's input $u_0$ can only be known by its neighbors, the proposed controller involves a local estimate of $u_0$. {\color{blue}Similarly, it also entails a local estimate of the leader's position $x_0$}. {\color{blue}Hence, an input observer is designed, which can be used by a follower to infer the leader's input.} Building on this input observer, another observer will be proposed for a follower to locally reconstruct the leader's position $x_0$. The design will be complete when the observers are integrated into the proposed controller. This methodology is illustrated in Figure~\ref{diagram-V1}.


To begin with, we consider the following control law for follower $i$:
\begin{align}\label{controller-first}
u_i=-k_{1}(x_i-\hat{x}_{0,i})+\hat{u}_{0,i},
\end{align}
where $k_1>0$ is the control gain, and $\hat{x}_{0,i}$ and $\hat{u}_{0,i}$ are follower $i$'s estimates of the leader's position and input, respectively. Here, the term $x_i-\hat{x}_{0,i}$ is meant to drive the follower approaching and tracking the leader, and the term $\hat{u}_{0,i}$  to ensure that the follower applies maneuvers consistent with the leader's driving input.

Proceeding further, we propose the following input observer for follower $i$ to estimate the leader's input $u_0$:
\begin{subequations}\label{u0-estimation-first}
\begin{align}\nonumber
\dot{z}_{i}=&-b_{i}lz_{i}-b_{i}^{2}l^{2}x_{0}-\sum_{j \in \mathcal{N}_{i}}a_{ij}(\hat{u}_{0,i}-\hat{u}_{0,j})\\
&-d_i \cdot\mathrm{sgn}\left[\sum_{j \in \mathcal{N}_{i}}a_{ij}(\hat{u}_{0,i}-\hat{u}_{0,j})+lb_{i}(\hat{u}_{0,i}-u_{0})\right], \\
\hat{u}_{0,i}=&z_{i}+b_{i}lx_{0}, \\ \label{adaptive-gain}
\dot{d}_i=&\tau_{i}\left|\sum_{j \in \mathcal{N}_{i}}a_{ij}(\hat{u}_{0,i}-\hat{u}_{0,j})+lb_{i}(\hat{u}_{0,i}-u_{0})\right|,
\end{align}
\end{subequations}
where $z_{i}$ is the observer's internal state, $l$ a scalar gain, $d_i$ an adaptive gain and $\tau_i$ is a positive scalar. This design is inspired by an unknown disturbance observer  developed in~\citep{yang:2013:ITIE}. However, we introduce two significant modifications. {\color{blue}First, the original design  in~\citep{yang:2013:ITIE} is a centralized observer for a single plant, whereas in this case it has been transformed to achieve distributed input estimation among a group of agents.} Second, {\color{blue}an adaptive mechanism is developed to enable a dynamic adjustment for the gain $d_i$,} as shown in~\eqref{adaptive-gain}, which helps avoid the cumbersome or inefficient gain selection procedure that would be necessary otherwise.

Building on the estimation of $u_0$ through~\eqref{u0-estimation-first}, a position observer is designed as follows:
\begin{align}\label{postion-filter-first}
\dot{\hat{x}}_{0,i}=-c\left[\sum_{j \in \mathcal{N}_{i}}a_{ij}(\hat{x}_{0,i}-\hat{x}_{0,j})
+b_{i}(\hat{x}_{0,i}-x_{0})\right]+\hat{u}_{0,i},
\end{align}
where $c$ is a scalar gain.
Note that the term $-\sum_{j \in \mathcal{N}_{i}}a_{ij}(\hat{x}_{0,i}-\hat{x}_{0,j}) -b_{i}(\hat{x}_{0,i}-x_{0})$ can help the observer overcome the error of the initial guess using neighborhood position estimation difference. The term $\hat{u}_{0,i}$ is to ensure that the observer's input is consistent with the leader's actual input $u_0$. With such a design, it is anticipated that $\hat{x}_{0,i}$ can converge to $x_0$.

Combining~\eqref{controller-first}-\eqref{postion-filter-first}, we obtain a complete description of an input-observer-based controller. Next, we will prove its convergence.

\subsection{Convergence Analysis}
To analyze its convergence properties, the next assumption and lemmas are needed.
\begin{assumption}\label{u0-constraint}
The input $u_0\in \mathcal{C}^1$, and its first-order derivative is bounded and satisfies $|\dot{u_0}|\leq w < \infty$, where $  w$ is unknown.
\end{assumption}
This assumption is mild and reasonable, since the leader's  maneuver input $u_0$ should be smooth and  bounded in rate-of-change due to practical control actuation limits. In addition, we assume that the bound for the rate-of-change does not have to be known. This reduces the amount of  information about the leader that must be available to followers. It may also help avoid potential conservatism in control design caused by a bound set too large.


Define $e_{u,i}=\hat{u}_{0,i}-u_0$, which is the input estimation error. According to~\eqref{u0-estimation-first}, the closed-loop dynamics of $e_{u,i}$ can be written as
\begin{align}\label{u0-dynamics-first}\nonumber
\dot{e}_{u,i}=&\dot{\hat{u}}_{0,i}-\dot{u}_0=\dot{z}_i+b_il\dot{x}_0-\dot u_0\\ \nonumber
=&-b_ile_{u,i}-\sum_{j \in \mathcal{N}_{i}}a_{ij}(\hat{u}_{0,i}-\hat{u}_{0,j})-\dot{u}_0\\
&-d_i\cdot\mathrm{sgn}\left[\sum_{j \in \mathcal{N}_{i}}a_{ij}(\hat{u}_{0,i}-\hat{u}_{0,j})+lb_{i}(\hat{u}_{0,i}-u_{0})\right].
\end{align}
Let us define $e_u=\left[\begin{matrix} e_{u,1}& e_{u,2} &\cdots & e_{u,N}\end{matrix}\right]^{\top}$. It then follows from~\eqref{u0-dynamics-first} that
\begin{align}\label{eu-dynamics-first}
\dot{e}_{u}=-H_1e_u-D\cdot\mathrm{sgn}(H_1e_u)-\dot{u}_0\mathbf{1},
\end{align}
where $H_1=lB+L$ and $D=\mathrm{diag}(d_{1},\ldots, d_{N})$. The convergence of $e_u$ to zero is shown in the following lemma.

\begin{lemma}\label{u0-analysis}
If Assumption~\ref{u0-constraint} holds, the input estimation $\hat{u}_{0,i}$ of~\eqref{u0-estimation-first} can track the input $u_0$ asymptotically with $\lim_{t\rightarrow\infty} e_u = 0$.
\end{lemma}

{\noindent\bf Proof:}
By Lemmas~\ref{L-notation} and~\ref{H-notation}, $H_1$ is positive definite. Consider the Lyapunov function ${\color{blue}V(e_u, d_i)}=\frac{1}{2}e_u^\top H_1e_u+\sum_{i=1}^N \frac{(d_i-\beta)^{2}}{2\tau_i}$ for the input estimation error dynamics in~\eqref{eu-dynamics-first}, where $\beta$ is a positive constant. The derivative of ${\color{blue}V(e_u, d_i)}$ is given by
\begin{align}\label{V-derivative-first}\nonumber
\dot{V}&=-e_u^\top H_1^2e_u-e_u^\top H_1 D\cdot \mathrm{sgn}(H_1e_u)-e_u^\top H_1\dot{u}_0\mathbf{1} +\sum_{i=1}^N \frac{(d_i-\beta)\dot{d_i}}{\tau_i}\\ \nonumber
&\leq -\sum_{i=1}^N d_i\left(\sum_{j \in \mathcal{N}_{i}}a_{ij}(\hat{u}_{0,i}-\hat{u}_{0,j})+lb_{i}(\hat{u}_{0,i}-u_{0})\right)^\top\\ \nonumber
&\quad \cdot\mathrm{sgn}\left(\sum_{j \in \mathcal{N}_{i}}a_{ij}(\hat{u}_{0,i}-\hat{u}_{0,j})+lb_{i}(\hat{u}_{0,i}-u_{0})\right)-e_u^\top H_1^2e_u+\sum_{i=1}^N \frac{(d_i-\beta)\dot{d_i}}{\tau_i}+w\|H_1e_u\|_1\\ \nonumber
&=-\sum_{i=1}^N d_i\left|\sum_{j \in \mathcal{N}_{i}}a_{ij}(\hat{u}_{0,i}-\hat{u}_{0,j})+lb_{i}(\hat{u}_{0,i}-u_{0})\right|
-e_u^\top H_1^2e_u+w\|H_1e_u\|_1 \\ \nonumber
&\quad +\sum_{i=1}^N (d_i-\beta)\left|\sum_{j \in \mathcal{N}_{i}}a_{ij}(\hat{u}_{0,i}-\hat{u}_{0,j})+b_{i}(\hat{u}_{0,i}-u_{0})\right| \\
&=-e_u^\top H_1^2e_u-(\beta-w)\|H_1e_u\|_1.
\end{align}
It is noted that $e_u^\top H_1^2e_u\geq 0$. Then, given $ w$, there always exists a $\beta$ that guarantees $\beta \geq w$. So we can obtain $\dot{V}\leq 0$ from~\eqref{V-derivative-first}, which indicates that ${\color{blue}V(e_u, d_i)}$ is non-increasing.
Therefore, one can see from the Lyapunov function that $e_u$ and $d_i$ are bounded.
By noting that $\tau_i>0$, it follows from~\eqref{adaptive-gain} that $d_i$ is monotonically increasing. Thus, the boundedness of $d_i$ indicates that each $d_i$ converges to some finite value. In the meantime, ${\color{blue}V(e_u, d_i)}$ reaches a finite limit as it is decreasing and lower-bounded by zero. Let us define $s(t) =\int_0^t e_u^\top (\tau)H_1^2e_u(\tau) d \tau$. It is obtained that $s(t) \leq V(0)-V(t) $ by integrating $\dot{V}\leq -e_u^\top H_1^2e_u$. Thus, $\lim_{t\rightarrow \infty} s(t)$ exists and is finite. Due to the boundedness of $e_u$ and $\dot e_u$, $\ddot s$ is also bounded. This shows that $\dot s$ is uniformly continuous. Hence, $\lim_{t\rightarrow \infty} \dot s(t) = 0$
{\color{blue}by Barbalat's} Lemma~\citep{khalil:1996:PHNJ}, indicating that $\lim_{t \rightarrow \infty}e_u=0$.
It is noted that~\eqref{u0-dynamics-first} is globally asymptotically stable. \hfill$\blacksquare$

Lemma~\ref{u0-analysis} indicates that each follower can successfully estimate the control input $u_0$ with the proposed input observer. We will next analyze the asymptotic stability of the position observer. The input-to-state stability lemma will be used.

\begin{lemma}~\citep{khalil:1996:PHNJ}~\label{ISS}
Consider an input-to-state stable (ISS) nonlinear system $\dot{x}=F(x,w)$. If the input satisfies $\lim_{t\rightarrow \infty}w(t)=0$, then the state $\lim_{t\rightarrow \infty}x(t)=0$.
\end{lemma}

Define follower $i$'s position estimation error as $e_{x,i}=\hat{x}_{0,i}-x_0$. The error vector for all followers is denoted as $e_x=\left[\begin{matrix} e_{x,1}& e_{x,2}& \cdots&e_{x,N} \end{matrix}\right]^{\top}$. It can be derived from~\eqref{postion-filter-first} that
\begin{align}\label{ex-dynamics-first}
\dot{e}_{x}=-cH_2e_x+e_u,
\end{align}
where $H_2=B+L$.

\begin{lemma}\label{x0-analysis}
If Assumption~\ref{u0-constraint} holds, the system in~\eqref{ex-dynamics-first} is asymptotically stable with $\lim_{t\rightarrow\infty}e_x = 0$, if the observer gain $c$ is chosen such that $c>0$.
\end{lemma}

{\noindent\bf Proof:}
According to Lemmas~\ref{L-notation} and~\ref{H-notation}, $H_2$ is positive definite. Then, the system in~\eqref{ex-dynamics-first} is ISS, and as a result, $\lim_{t\rightarrow \infty}e_x=0$ holds.
\hfill$\blacksquare$

The above lemma shows the effectiveness of the proposed position observer for a follower to estimate the leader's  $x_0$. Now, let us prove the convergence of the tracking control.
Define follower $i$'s tracking error as $e_i=x_i-x_0$, and put together $e_i$ for $i=1,2,\ldots,N$ to form the vector $e=\left[\begin{matrix} e_{1}& e_{2} &\cdots & e_{N}\end{matrix}\right]^{\top}$. Using~\eqref{follower-dynamics-first} and~\eqref{controller-first}, it can be derived that the dynamics of $e$ is governed by 
\begin{align}\label{e-dynamics-first}
\dot{e}=-k_{1}e+k_{1}e_x+e_u.
\end{align}
The theorem below shows that $e$ will approach 0 as $t\rightarrow \infty$.

\begin{theorem}\label{ui-analysis-first}
Suppose that Assumption~\ref{u0-constraint} is satisfied. If the observer gain is chosen such that $k_1>0$ holds, the system in~\eqref{e-dynamics-first} is asymptotically stable, and $\lim_{t\rightarrow \infty} |x_i(t)-x_0(t)| =0$ for $i=1,2,\ldots,N$.
\end{theorem}

{\noindent\bf Proof:}
It can be obtained from Lemma~\ref{ISS} that the system in~\eqref{e-dynamics-first} is ISS if $k_1>0$. Therefore, $\lim_{t\rightarrow \infty}e=0$ results from the analysis in Lemmas~\ref{u0-analysis} and~\ref{x0-analysis}. This completes the proof.
\hfill$\blacksquare$

Theorem~\ref{ui-analysis-first} shows that the proposed tracking control approach would enable  each follower to approach and track the leader as time goes by, with the position tracking error converging to zero. The following remark further summarizes its difference from some existing methods and advantages.

\begin{remark}
The input-observer-based tracking control approach proposed above presents a few advantages over many existing methods. First, for this approach, a follower only needs to interchange information with its neighbors. By comparison,  some studies in the literature requires that the leader's input must be known by any follower even if it is not a neighbor of the leader, e.g.,~\citep{hong:2008:distributed,li:2010:ITCS,yu:2016:IJAS,hu:2010:AUT}. Second, the followers do not have to be given information about the leader's controller. This contrasts with~\citep{Wang:CTT:2016}, which stipulates that every follower knows the leader's exact control law, and with~\citep{cao:2012:TAC}, which requires the upper bound of the leader's control input to be known by all followers. Finally, the approach relaxes the assumption about the leader's control input. Here, a bound is only imposed on its rate-of-change rather than its magnitude as in~\citep{lihua:2013:TAC}. This implies that this approach can apply to the case when the leader applies high-magnitude maneuvers. In particular, the bound of rate-of-change does not have to be known for the control design, further conducive to practical application of the proposed approach. \hfill$\bullet$
\end{remark}

\subsection{Extension to a Simplified Case}

A general case is considered above that the leader's input $u_0$ has a bounded rate-of-change. However, it is also practically meaningful in reality to consider a special case when the time derivative of $u_0$ becomes zero as time goes by. In other words, whatever the leader's movement is like at the beginning time, it gradually transitions to and maintains constant-speed movement. An example is a group of aerial vehicles tracking a leader that cruises at a stable speed to achieve high-quality photographing~\citep{smith:2016:RN}. This setting is also of considerable interest in the literature, e.g.,~\citep{zhao:2013:SCL}. Along this line, let us consider that the rate-of-change of $u_0$ approaches zero, i.e., $\lim_{t\rightarrow \infty} \dot{u}_0(t)=0$. To deal with this case, we can reduce the input observer in~\eqref{u0-estimation-first} to the following form, which is structurally more concise:
\begin{subequations}\label{u0-estimation-zero-first}
\begin{align}
&\dot{z}_{i}=-b_{i}lz_{i}-b_{i}^{2}l^{2}x_{0}-\sum_{j \in \mathcal{N}_{i}}a_{ij}(\hat{u}_{0,i}-\hat{u}_{0,j}) , \\
&\hat{u}_{0,i}=z_{i}+b_{i}lx_{0}.
\end{align}
\end{subequations}
When this observer is integrated into the controller in~\eqref{controller-first}, effective tracking can be guaranteed under relaxed conditions. This argument is presented in the following corollary. The proof is straightforward and thus omitted here.

\begin{corollary}
Consider the systems in~\eqref{follower-dynamics-first} and assume that $\lim_{t\rightarrow \infty} \dot{u}_0(t)=0$. Suppose that the controller in~\eqref{controller-first} is applied together with the position observer in~\eqref{postion-filter-first} and input observer in~\eqref{u0-estimation-zero-first}. Then, $\lim_{t\rightarrow \infty} |x_i(t)-x_0(t)| =0$ for $i=1,2,\ldots,N$ if the control gain $k_1>0$ and the observer gain $c>0$.
\end{corollary}

\begin{remark}
In addition to structural conciseness, it is noted that this input observer does not require the leader's input information if compared to the one in~\eqref{u0-estimation-first}. This indicates that the leader does not even have to send its input to its neighbors in the considered setting, as a further advantage in practice. \hfill$\bullet$
\end{remark}

The above result indicates that one can potentially design different input observers within the proposed framework according to problem settings or practical needs. A further evidence is that an input observer designed in Section~\ref{second-order} can also be proven effective in achieving input estimation if applied here. 


\section{Second-order Leader-follower Tracking}\label{second-order}

This section considers leader-follower tracking for agents with second-order dynamics.
Now, the leader and followers are described as
\begin{align}\label{leader-follower-dynamics-second}
\left\{
\begin{array}{l}
\dot{x}_{i}=v_{i},\quad x_{i}\in \mathbb{R},\\
\dot{v}_{i}=u_{i}, \quad v_{i}\in \mathbb{R}, \ i = 0,1,\ldots,N,
\end{array}
\right.
\end{align}
where $x_{i}$ is the position, $v_{i}$ the velocity and $u_{i}$ the {\color{blue}input force}. Still, agent 0 is the leader, and the other agents numbered from $1$ to $N$ are followers. It is considered here that no velocity sensor is used by the leader and followers, i.e.,  $v_i$ for $i=0,1,\ldots,N$ is not measured.
Akin to the first-order case, our aim here is still to design a distributed control approach for each follower to track the leader, achieving $\lim_{t\rightarrow \infty}|x_{i}(t)-x_{0}(t)|=0$ and $\lim_{t\rightarrow \infty}|v_{i}(t)-v_{0}(t)|=0$.

\begin{figure}[t]
\begin{center}
\includegraphics[scale=0.39]{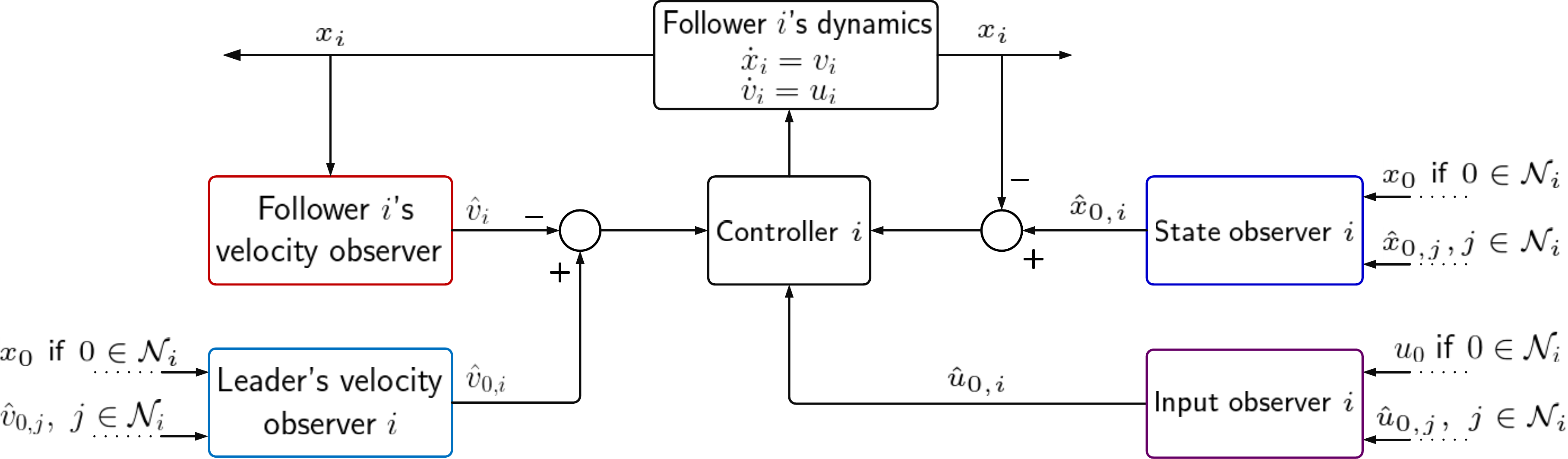}
 \caption{Input-observer-based framework for second-order leader-follower tracking.
}
\label{diagram-V2}
\end{center}
\end{figure}

To address this second-order tracking problem, we continue to leverage the design thinking of input-observer-based control. The specific design can be laid out in two main steps. First, a linear continuous tracking controller is proposed for a follower, which uses the follower's position measurement and a few estimates, including its own velocity and the leader's position, velocity and input. Second, a series of observers are progressively developed to obtain the needed estimates. An input observer is designed such that the follower can reconstruct the leader's input. This is followed by the development of two observers that permit it to estimate the leader's velocity and position, respectively. Another observer is also proposed to help the follower determine its own velocity. Combining these observers with the controller then enables tracking control. {\color{blue}This framework is schematically illustrated in Figure 2.}

Along the above line, we start with proposing a  control law for follower $i$, which is given by
\begin{align}\label{controller-second}
u_i=-k_{1}(x_i-\hat{x}_{0,i})-k_{2}(\hat{v}_i-\hat{v}_{0,i})+\hat{u}_{0,i},
\end{align}
where $k_1>0$ and $k_2>0$ are the controller gains. Here, $\hat{x}_{0,i}$, $\hat v_{0,i}$ and $\hat u_{0,i}$ are follower $i$'s estimates of the leader's position $x_0$, velocity $v_0$ and input $u_0$, and $\hat{v}_i$ represents agent $i$'s estimate of its own velocity $v_i$. Furthermore, the term $x_i-\hat{x}_{0,i}$ is used to propel follower $i$ to move toward the leader, and the term $\hat{v}_i-\hat{v}_{0,i}$ to synchronize its velocity with the leader's. The term $\hat u_{0,i}$ is intended to maintain follower's maneuver at the same level with the leader. Next, we build observers to obtain $\hat{u}_{0,i}$, $\hat{v}_i$, $\hat{v}_{0,i}$ and $\hat{x}_{0,i}$.

We firstly propose an input observer to estimate $u_0$ as follows:
\begin{subequations}\label{u0-estimation-second}
\begin{align}\nonumber
\dot{\hat{u}}_{0,i}=&-\sum_{j \in \mathcal{N}_{i}}a_{ij}(\hat{u}_{0,i}-\hat{u}_{0,j})-b_{i}
(\hat{u}_{0,i}-u_{0})\\
&-d_i\cdot\mathrm{sgn}\left[\sum_{j \in \mathcal{N}_{i}}a_{ij}
(\hat{u}_{0,i}-\hat{u}_{0,j})+b_{i}
(\hat{u}_{0,i}-u_{0})\right],\\
\dot{d}_i=&\tau_{i}\left|\sum_{j \in \mathcal{N}_{i}}a_{ij}(\hat{u}_{0,i}-\hat{u}_{0,j})+b_{i}(\hat{u}_{0,i}-u_{0})\right|.
\end{align}
\end{subequations}
Here, the term $-\sum_{j \in \mathcal{N}_{i}}a_{ij}(\hat{u}_{0,i}-\hat{u}_{0,j})
-b_{i}(\hat{u}_{0,i}-u_{0})$ is used to drive $\hat{u}_{0,i}$ toward approaching $u_0$; the $\mathrm{sgn}(\cdot)$ term is employed to maintain synchronization between $\hat u_{0,i}$ and $u_0$ in the presence of $\dot{u}_0$. It is seen that this observer does not require position $x_0$ measurement, differing from the one proposed earlier in~\eqref{u0-estimation-first}. Note that this input observer is also applicable to the first-order case with provable asymptotic stability. In other words, if it replaces~\eqref{u0-estimation-first}, the first-order tracking control can still be achieved under some mild conditions. {\color{blue} This implies that one can design different  kinds of   observers to achieve   estimation of the leader's input.} Then, $\hat{u}_{0,i}$ can be used to estimate $v_0$ using the observer
\begin{subequations}\label{v0-estimation}
\begin{align}
&\dot{z}_{i}=-b_{i}lz_{i}-b_{i}^{2}l^{2}x_{0}-\sum_{j \in \mathcal{N}_{i}}a_{ij}(\hat{v}_{0,i}-\hat{v}_{0,j})+\hat{u}_{0,i}, \\
&\hat{v}_{0,i}=z_{i}+b_{i}lx_{0},
\end{align}
\end{subequations}
where $z_{i}$, $l$, and $\hat{v}_{0,i}$ are the internal state of the observer, the observer gain, and the estimate of $v_0$, respectively. This velocity observer, as is seen, allows distributed estimation of the leader's velocity among all agents, even though it is not measured by a sensor. On such a basis, a position observer is designed for follower $i$ to estimate $x_0$:
\begin{align}\label{postion-filter-second}
\dot{\hat{x}}_{0,i}=- c\left[\sum_{j \in \mathcal{N}_{i}}a_{ij}(\hat{x}_{0,i}-\hat{x}_{0,j})
+b_{i}(\hat{x}_{0,i}-x_{0})\right]+\hat{v}_{0,i}.
\end{align}
Finally, follower $i$ uses the following observer to estimate its own velocity as it also has no velocity sensor:
\begin{align}\label{vi-estimation}
&\dot{\bar{z}}_{i}=-l\bar{z}_{i}-l^{2}x_{i}+u_i \nonumber\\
&\hat{v}_{i}=\bar{z}_{i}+lx_{i},
\end{align}
where $\bar{z}_{i}$ is the internal state of the observer.
Putting together the above observers~\eqref{u0-estimation-second}-\eqref{vi-estimation} with the controller~\eqref{controller-second}, we can obtain a tracking control approach. Its convergence will be analyzed next.
Yet before proceeding to the proof, we remark that Assumption~\ref{u0-constraint} is also needed here and for simplicity do not restate it.
In addition, the following lemmas will be used.
\begin{lemma}~\citep{kovacs:1999:AMM}\label{matrix-det}
Let $Q=\left[\begin{matrix}A&B\\C&D\end{matrix}\right]$, where $A$, $B$, $C$, $D\in \mathbb{R}^{n \times n}$. Then $\det(Q)=\det(AD-BC)$, if matrix $A$, $B$, $C$ and $D$ commute pairwise.
\end{lemma}

\begin{lemma}~\citep{yu:2011:AUT}\label{polynomial}
Given a complex coefficient polynomial of order two as follows:
\begin{align}
h(s)=s^2+(a_1+\mathbf{i}b_1)s+a_0+\mathbf{i}b_0,
\end{align}
where $\mathbf{i}=\sqrt{-1}$; $a_1$, $b_1$, $a_0$ and $b_0$ are real constraints. Then, $h(s)$ is stable if and only if $a_1>0$ and $a_1b_1b_0+a_1^2a_0-b_0^2>0$.
\end{lemma}

The following theorem is the main result regarding the convergence of the proposed tracking controller.
\begin{theorem}\label{ui-analysis-second}
Suppose that Assumption~\ref{u0-constraint} holds and apply the proposed control approach~\eqref{controller-second}-\eqref{vi-estimation} to the considered second-order systems in~\eqref{leader-follower-dynamics-second}. If $k_1>0$, $k_2>0$, $l >0$ and $c>0$, then $\lim_{t\rightarrow \infty}|x_{i}(t)-x_{0}(t)|=0$ and $\lim_{t\rightarrow \infty}|v_{i}(t)-v_{0}(t)|=0$.
\end{theorem}

{\noindent\bf Proof:}
It can be derived from~\eqref{u0-estimation-second} that the dynamics of the input estimation error $e_u$ is given by
\begin{align}\label{u0-dynamics-second}
\dot{e}_{u}=-H_2e_u- D \cdot \mathrm{sgn}(H_2e_u)-\dot{u}_0\mathbf{1}.
\end{align}
Along similar lines to the proof of Lemma~\ref{u0-analysis}, the above system is asymptotically stable, i.e., $\lim_{t\rightarrow \infty} e_u=0$.

Define the velocity estimation error $e_{0v,i}$ as $e_{0v,i}=\hat{v}_{0,i}-v_0$. According to~\eqref{v0-estimation}, the dynamics of $e_{0v,i}$ can be written as
\begin{align}\label{v0-dynamics}
\dot{e}_{0v,i}&=\dot{\hat{v}}_{0,i}-\dot{v}_0
=-b_ile_{0v,i}-\sum_{j \in \mathcal{N}_{i}}a_{ij}(\hat{v}_{0,i}-\hat{v}_{0,j})+\hat{u}_{0,i}-u_0.
\end{align}
Further, let us define the vector $e_{0v}=\left[\begin{matrix} e_{0v,1}& e_{0v,2}&\cdots& e_{0v,N}\end{matrix}\right]^{\top}$. The dynamics of $e_{0v}$ then can be obtained from~\eqref{v0-dynamics}, which is
\begin{align}\label{e0v-dynamics}
\dot{e}_{0v}=-H_1e_{0v}+e_u.
\end{align}
Because of $\lim_{t\rightarrow \infty} e_u=0$ and the ISS result in Lemma~\ref{ISS}, it can be concluded that $\lim_{t\rightarrow \infty} e_{0v}=0$.

By~\eqref{postion-filter-second}, the position estimation error vector $e_x$, which shares the same definition as in the first-order case, is governed by the following dynamics equation:
\begin{align}\label{ex-dynamics-second}
\dot{e}_{x}=-cH_2e_{x}+e_{0v}.
\end{align}
According to Lemma~\ref{ISS}, the system in~\eqref{ex-dynamics-second} is ISS. Since $\lim_{t\rightarrow \infty} e_{0v}=0$, we have  $\lim_{t\rightarrow \infty} e_{x}=0$.

Now we consider a follower's estimation error for its own velocity. Denote $e_{v,i}=\hat{v}_i-v_i$ and $e_{v}=\left[\begin{matrix} e_{v,1}&e_{v,2}&\cdots&e_{v,N}\end{matrix}\right]^{\top}$. We can derive the dynamics of $e_v$ from~\eqref{vi-estimation}, which is
\begin{align}\label{ev-dynamics}
\dot{e}_{v}=-le_{v}.
\end{align}
Obviously, $\lim_{t\rightarrow \infty} e_{v}=0$ if $l>0$.

Consider the leader and followers in~\eqref{leader-follower-dynamics-second} under the control law~\eqref{controller-second}, one can obtain follower's closed-loop dynamics:
\begin{subequations}\label{closed-loop-second}
\begin{align}
\dot{x}_{i}-\dot{x}_{0}=&v_i-v_{0},\\ \nonumber
\dot{v}_{i}-\dot{v}_{0}=&-k_{1}(x_i-x_0)-k_{2}(v_i-v_{0})-k_{2}(\hat{v}_{i}-v_i)\\
&+k_{2}(\hat{v}_{0,i}-v_0)         +k_{1}(\hat{x}_{0,i}-x_0)
+\hat{u}_{0,i}-u_{0},
\end{align}
\end{subequations}
for $i=1,2,\ldots,N$.
Define $e=\left[\begin{matrix} x_1-x_0&\cdots&x_N-x_0 & v_1-v_0& \cdots & v_N-v_0 \end{matrix}\right]^{\top}$. Then, combining~\eqref{u0-dynamics-second}, \eqref{ev-dynamics} and~\eqref{closed-loop-second}, we have the closed-loop tracking error dynamics of the entire leader-follower system:
\begin{align}\label{e-dynamics-second}
\dot{e}=F_1e+F_2,
\end{align}
where
\begin{align*}
F_1&=\left[\begin{matrix}0&I\\-k_1I&-k_2I\end{matrix}\right],\
F_2 =\left[\begin{matrix}0\\-k_2e_v+k_2e_{0v}+k_1e_x+e_u\end{matrix}\right].
\end{align*}
Furthermore, according to Lemma~\ref{matrix-det}, the characteristic polynomial of $F_1$ is given by
\begin{align}\label{det-F}\nonumber
\det(sI-F_1)&=\det\left(\left[\begin{matrix} sI&-I\\k_1I&sI+k_2I   \end{matrix}\right]\right)\\\nonumber
&=\det(s^2I+k_2sI+k_1I)\\
&=\prod_{i=1}^{N}(s^2+k_2s+k_1)=\prod_{i=1}^{N}h_i(s).
\end{align}
Based on Lemma~\ref{polynomial}, $h_i(s)$ is stable when $k_1>0$ and $k_2>0$. With this result, the system~\eqref{e-dynamics-second} is ISS as $\lim_{t\rightarrow \infty} F_2=0$ from~\eqref{u0-dynamics-second}-\eqref{ev-dynamics}. Hence, $\lim_{t\rightarrow \infty} e=0$, which implies $\lim_{t\rightarrow \infty}|x_{i}(t)-x_{0}(t)|=0$ and $\lim_{t\rightarrow \infty}|v_{i}(t)-v_{0}(t)|=0$. This completes the proof.
\hfill$\blacksquare$

\begin{remark}
This proposed tracking control approach offers some merits when compared with the literature. First, it does not require a follower to know the leader's input or velocity if they are not neighbors, differing from~\citep{hong:2008:distributed,zhu:2010:AUT,cao:2012:TAC,hu:2010:AUT}. This is similar to the approach in Section~\ref{first-order} and attributed to the input and velocity observers giving a follower a crucial ``leader-awareness''. Second, this approach can enable accurate tracking in the absence of velocity sensors. Recent years have seen a growing interest in tracking control without velocity measurements due to its practical benefits. Our proposed approach is different from the present methods in some interesting ways. Through the velocity observers, it makes an explicit estimation of the leader's and follower's velocities. This differs from~\citep{Ghapani:AUTO:2017,Zhang:SCL:2014,Zhou:CTA:2014}, which make no velocity estimation and use only neighborhood position difference to achieve velocity-free tracking control. Velocity observer design is also considered in~\citep{hu:2010:AUT}. However, the design therein requires the leader's {\color{blue}input force} to be known by every follower. By contrast, our approach obviates this need because  the input observer can infer the leader's input. \hfill$\bullet$
\end{remark}

\begin{figure}[t]
\begin{center}
\includegraphics[scale=0.4]{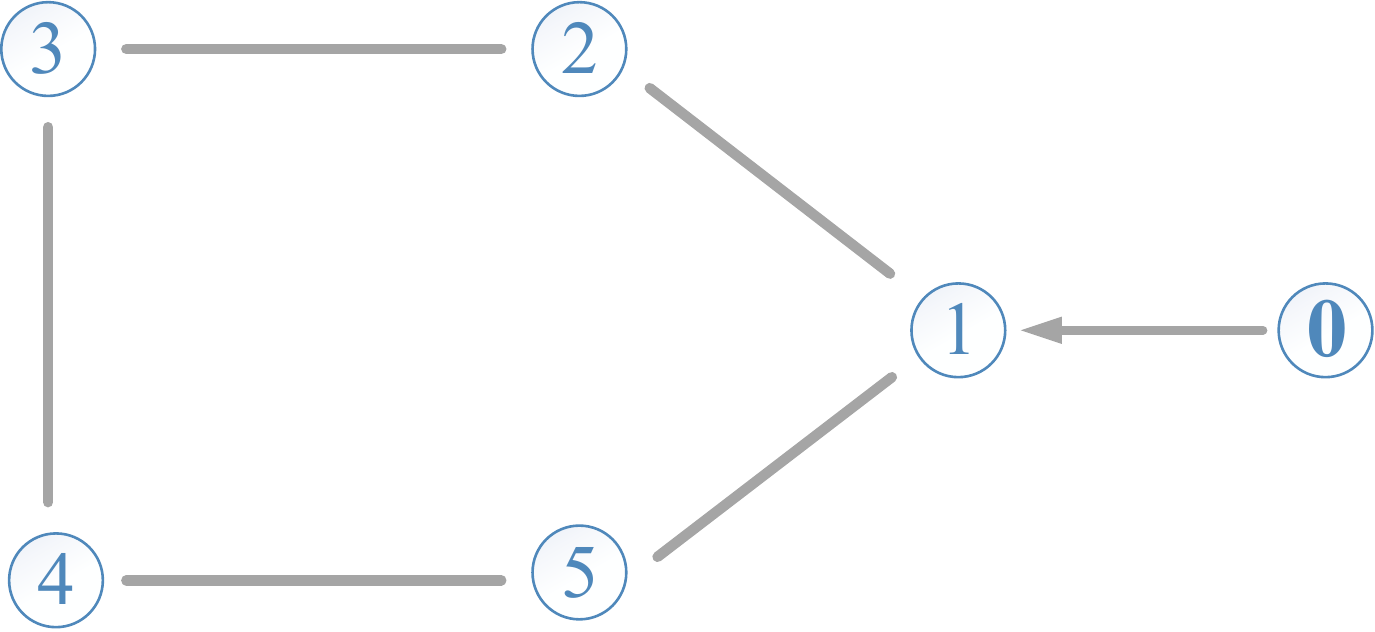}
 \caption{Communication topology of the MAS in simulation.
}
\label{topology}
\end{center}
\end{figure}

\begin{figure*}[t]
\normalsize
\centering
\subfloat[]{
\includegraphics[trim={7mm 1mm 11mm 7mm},clip,width=0.38\linewidth]{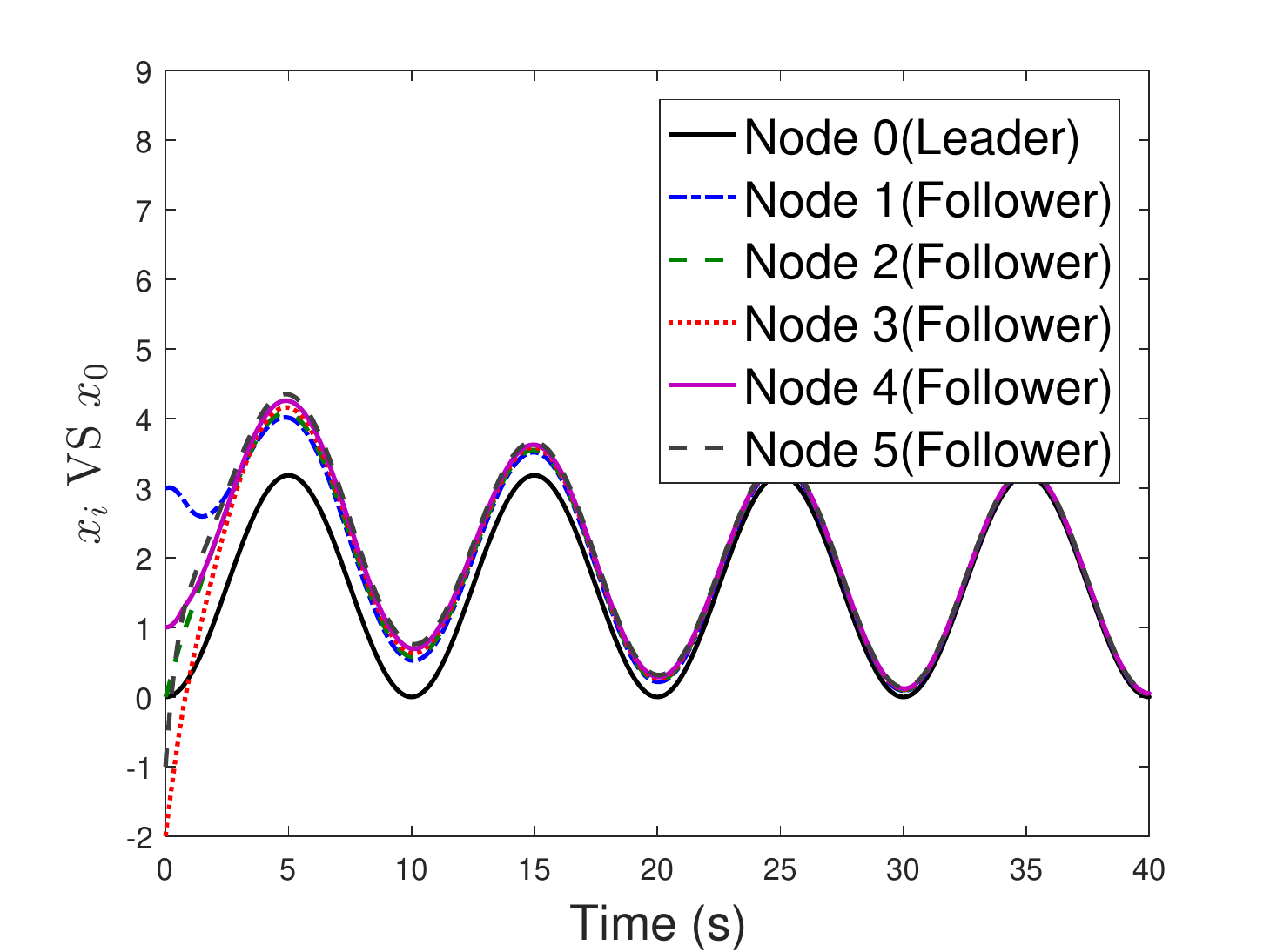}\label{position-first}} \hspace{8mm}
\subfloat[]{
\includegraphics[trim={5mm 1mm 12mm
 6mm},clip,width=0.38\linewidth]{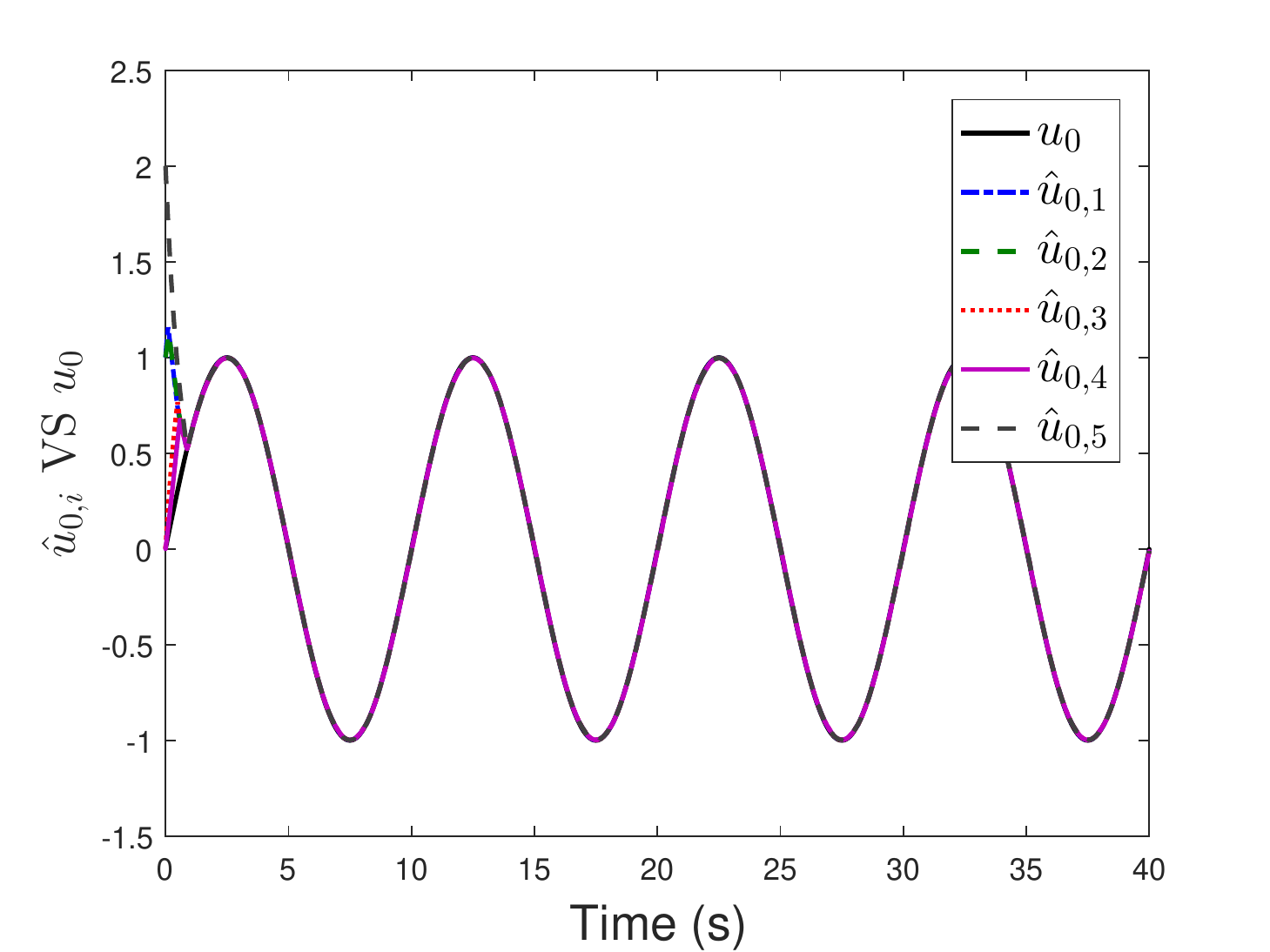}\label{u0-first}}\\
\subfloat[]{
\includegraphics[trim={5mm 1mm 11mm
7mm},clip,width=0.38\linewidth]{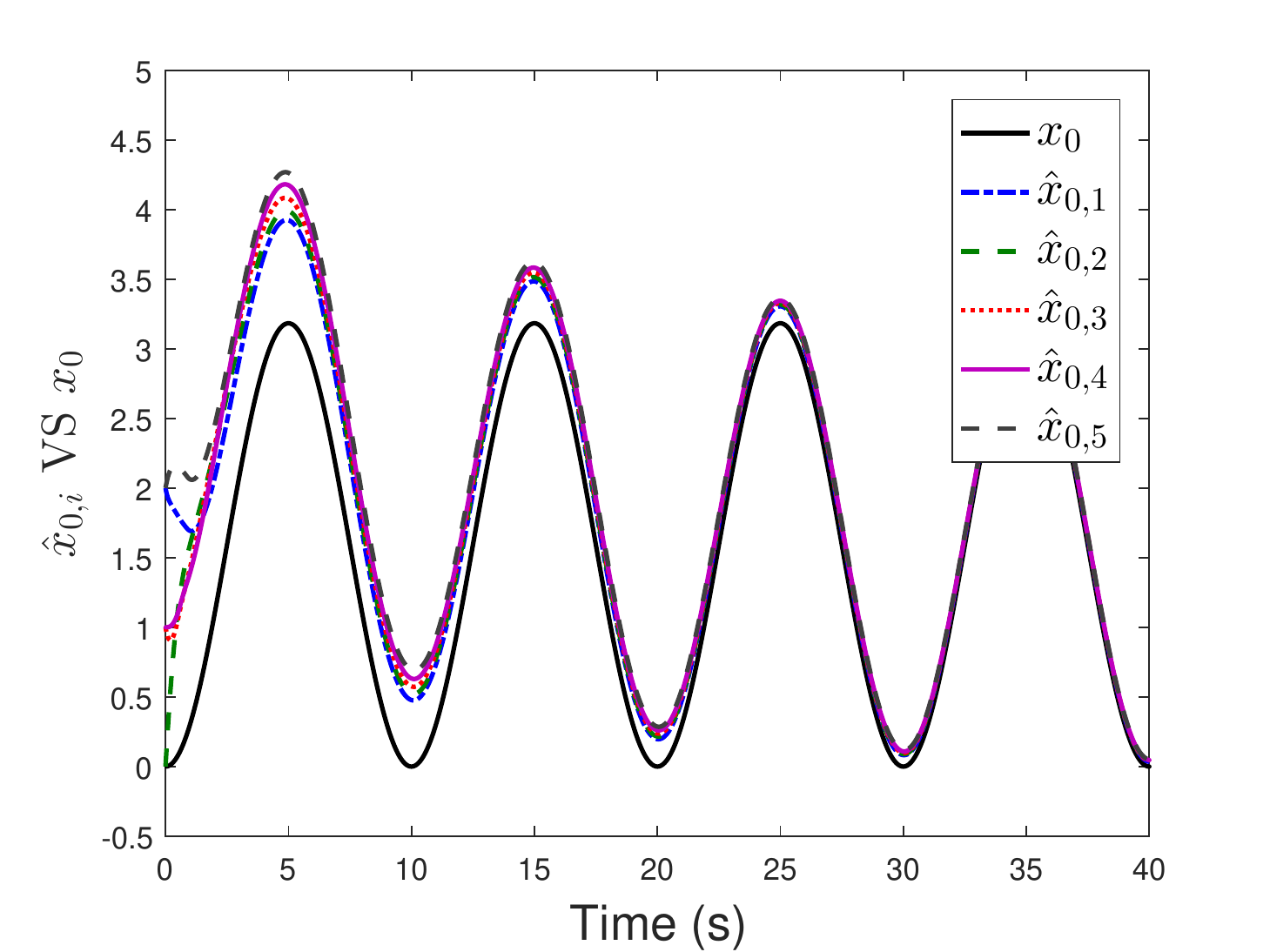}\label{x0-first}}\hspace{8mm}
\subfloat[]{
\includegraphics[trim={5mm 1mm 12mm
 6mm},clip,width=0.38\linewidth]{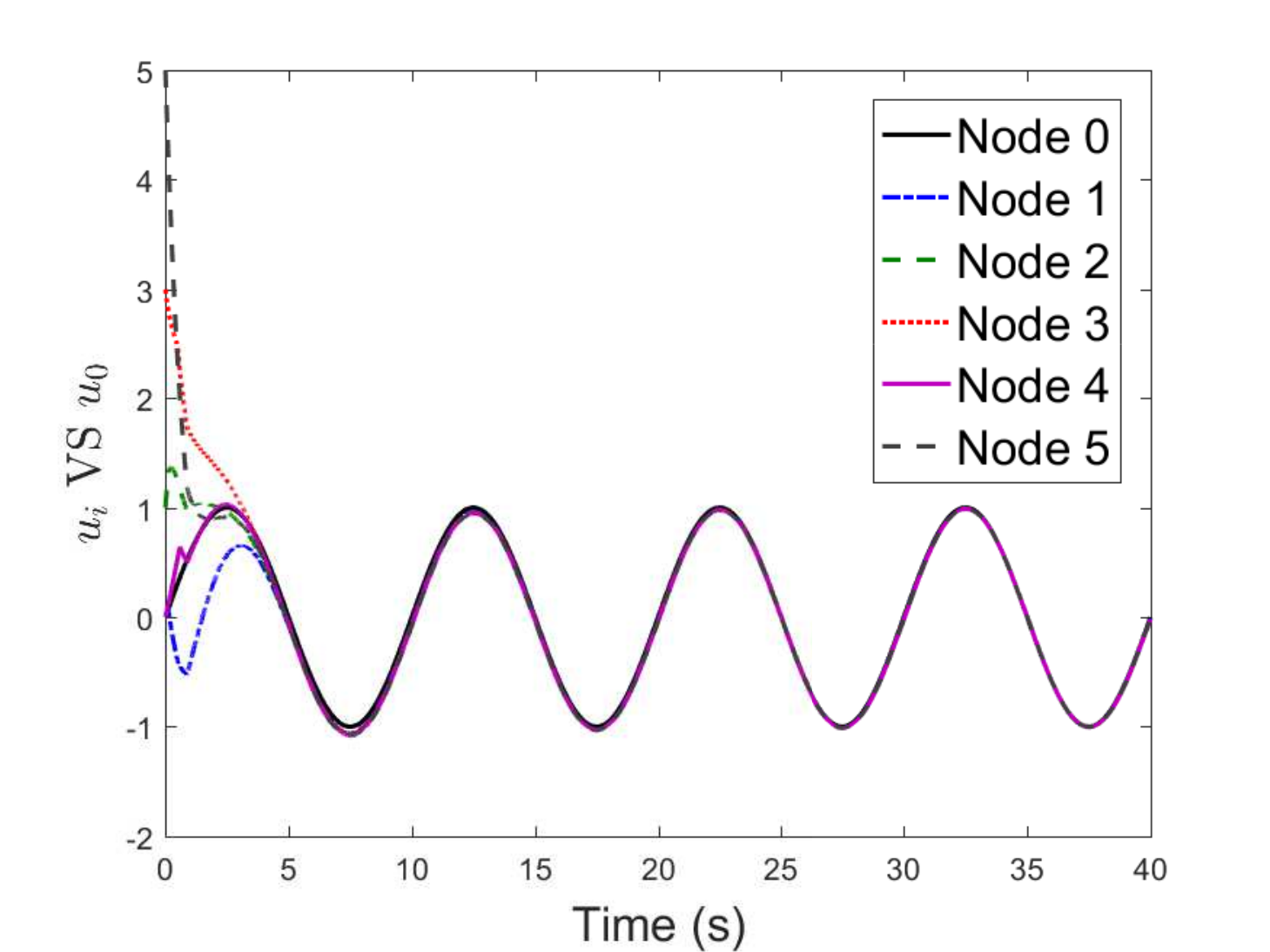}\label{ui-first}}
\caption{Tracking control for  a first-order MAS: (a)  position tracking; (b) followers' estimation of the leader's input; (c) followers' estimation of the leader's position; (d)  follower's input profiles in comparison with the leader's.}\label{First-order-MAS-profiles}
\end{figure*}

\begin{figure*}[t]
\normalsize
\centering
\subfloat[]{
\includegraphics[trim={6mm 1mm 11mm 7mm},clip,width=0.38\linewidth]{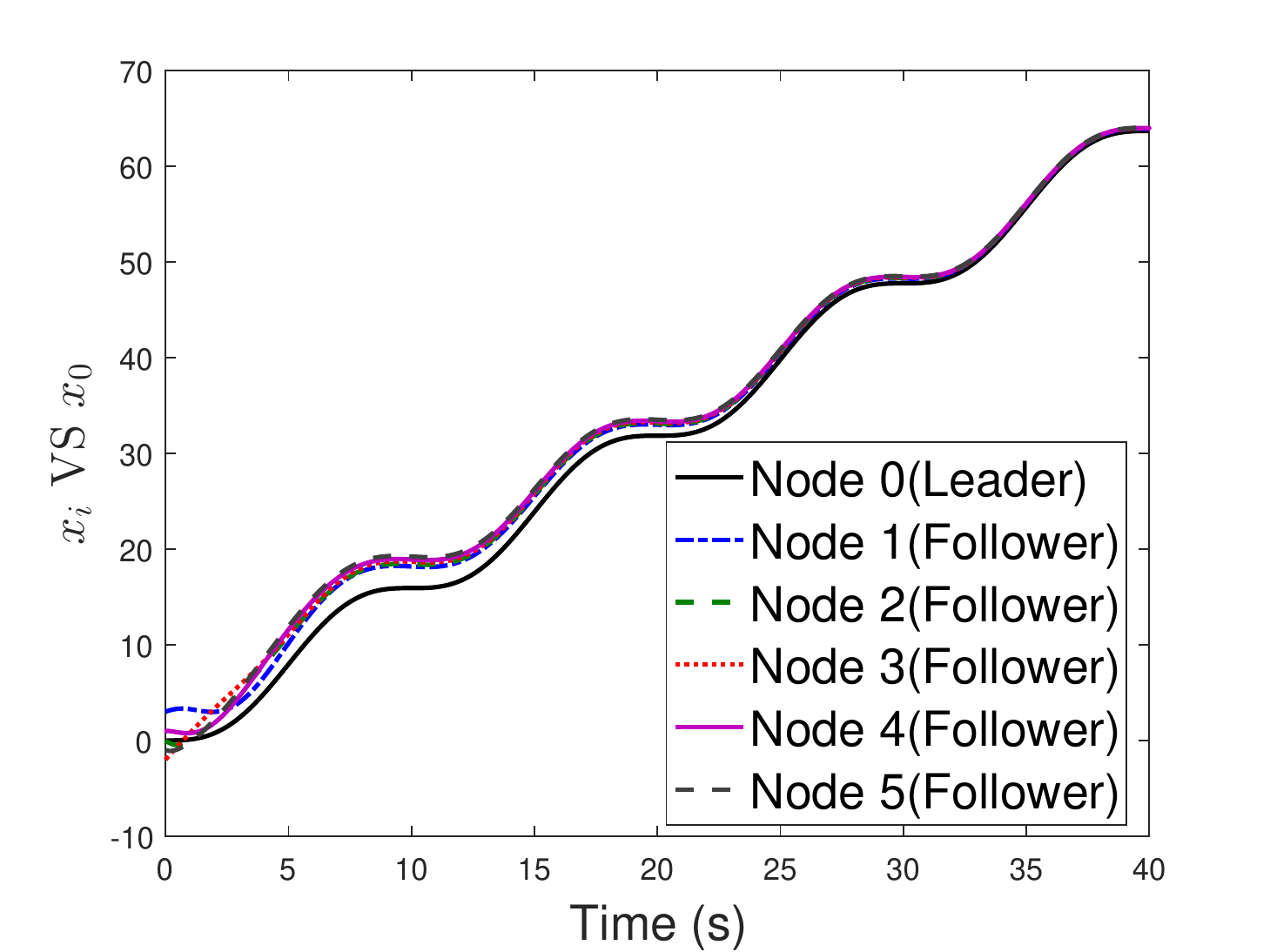}\label{position-second}}\hspace{8mm}
\subfloat[]{
\includegraphics[trim={5mm 1mm 12mm
 6mm},clip,width=0.38\linewidth]{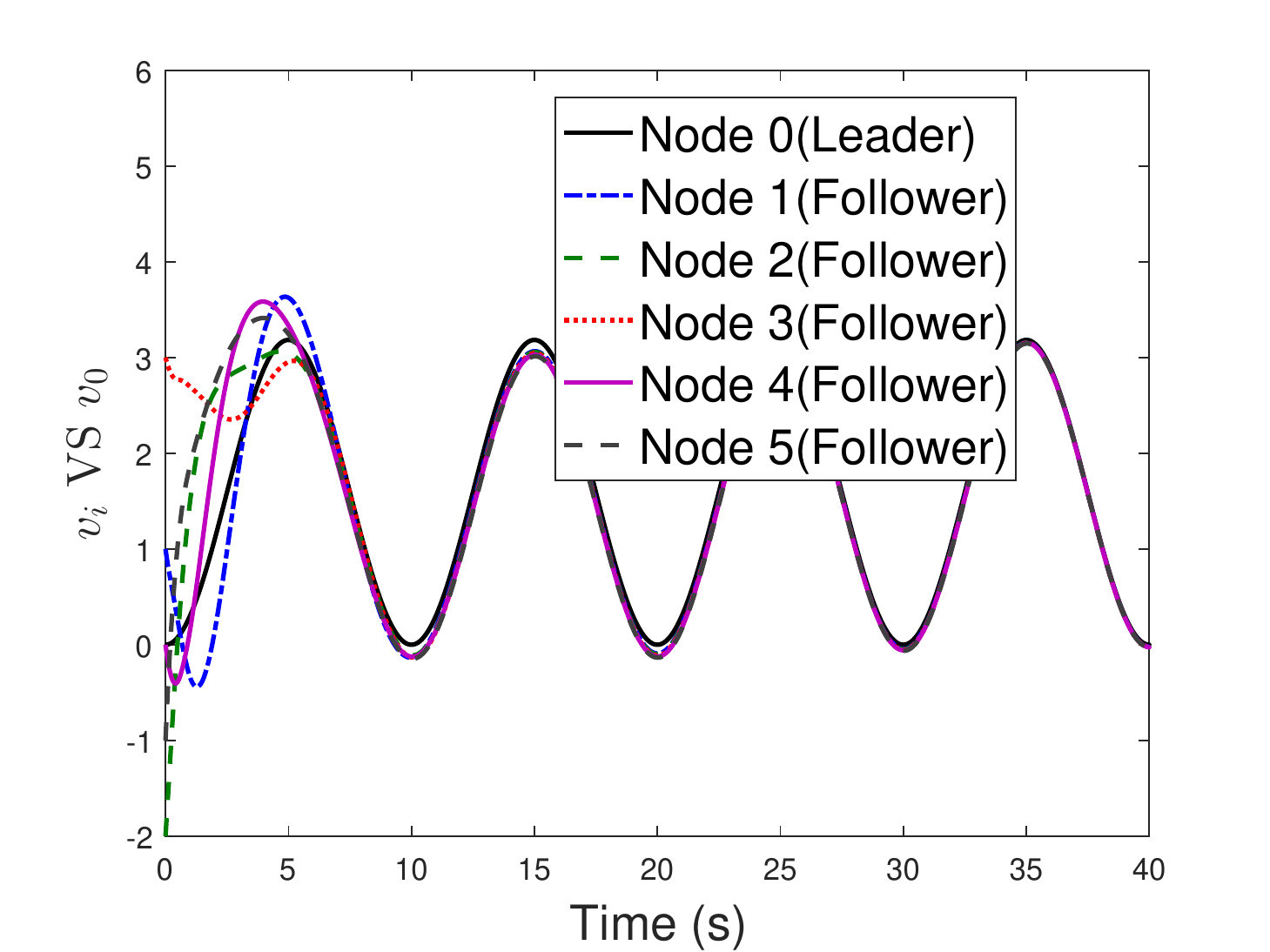}\label{velocity-second}}\\
 \subfloat[]{
\includegraphics[trim={6mm 1mm 11mm 7mm},clip,width=0.38\linewidth]{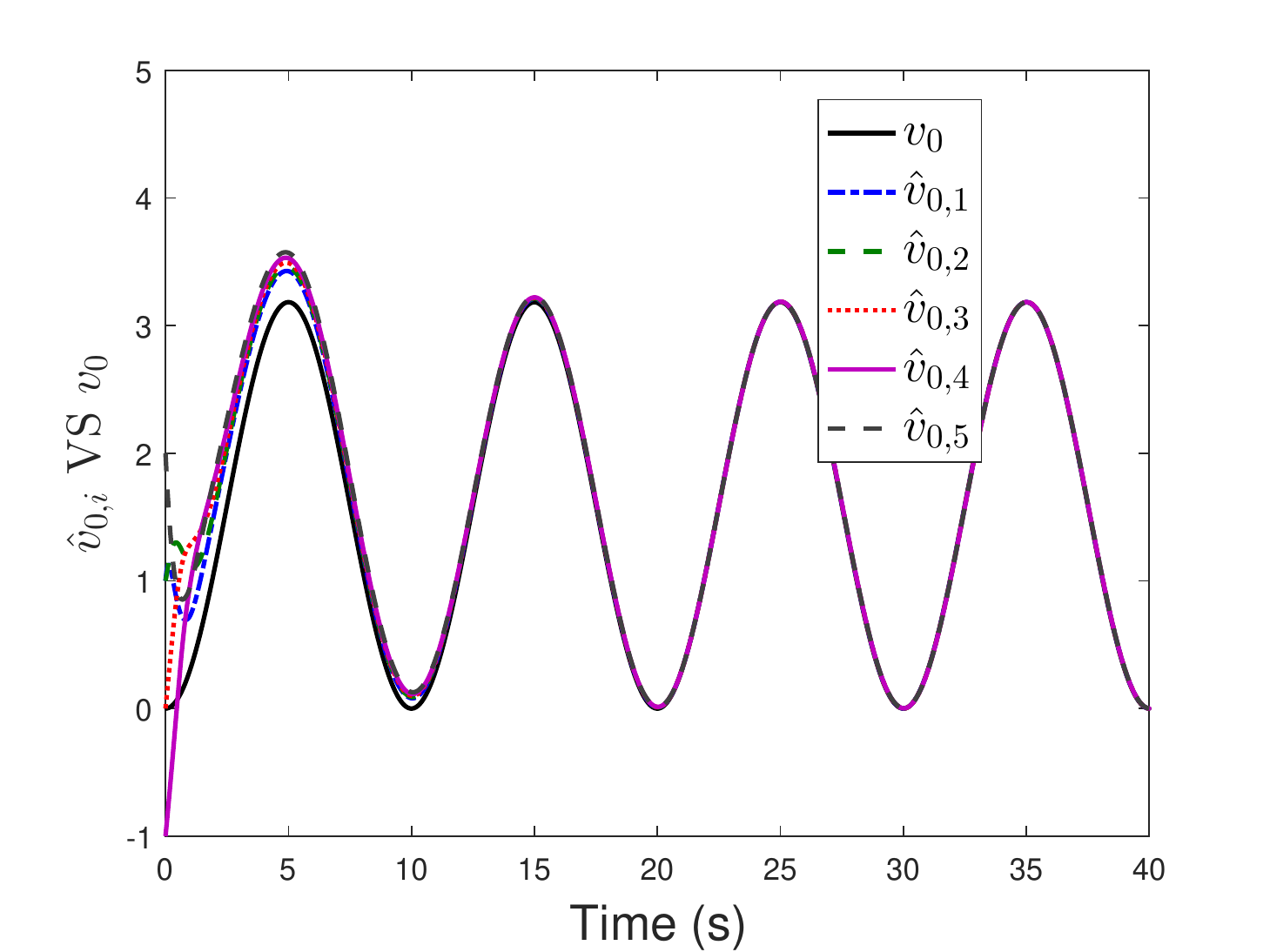}\label{leader-velocity-second}}\hspace{8mm}
\subfloat[]{
\includegraphics[trim={5mm 1mm 12mm
 6mm},clip,width=0.38\linewidth]{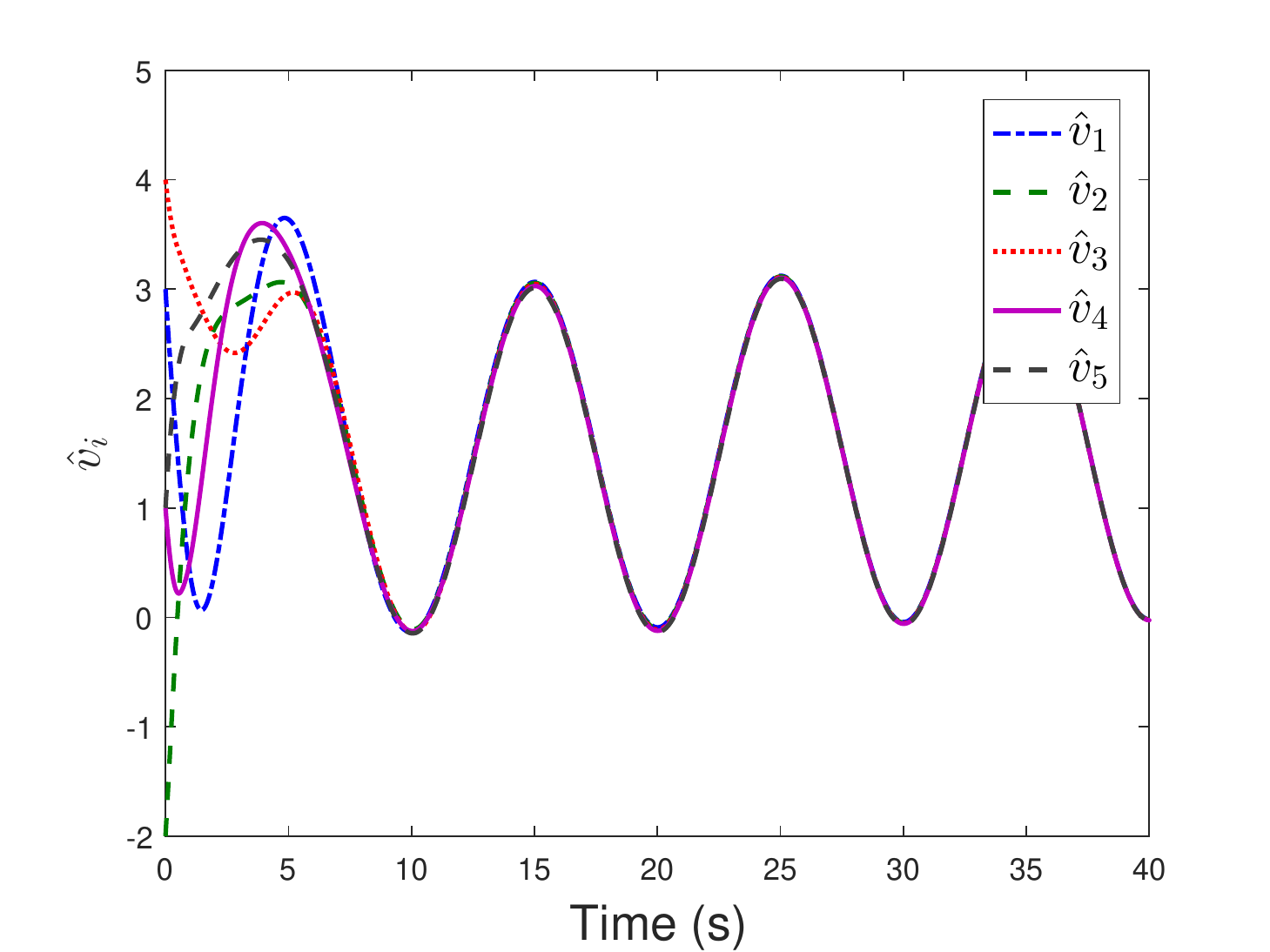}\label{follower-velocity-second}}\\
\subfloat[]{
\includegraphics[trim={5mm 1mm 11mm
7mm},clip,width=0.38\linewidth]{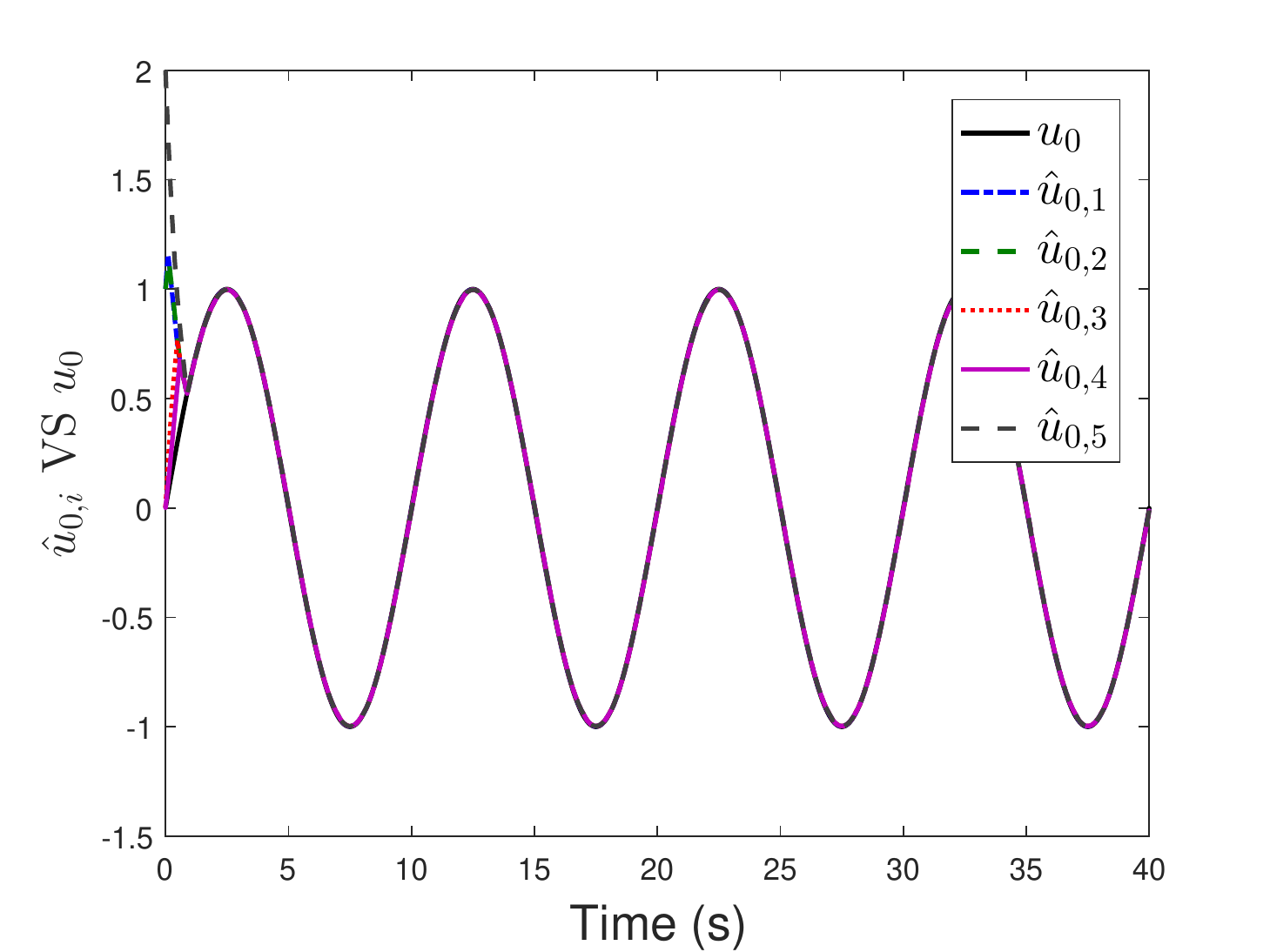}\label{u0-second}}\hspace{8mm}
\subfloat[]{
\includegraphics[trim={5mm 1mm 11mm
7mm},clip,width=0.38\linewidth]{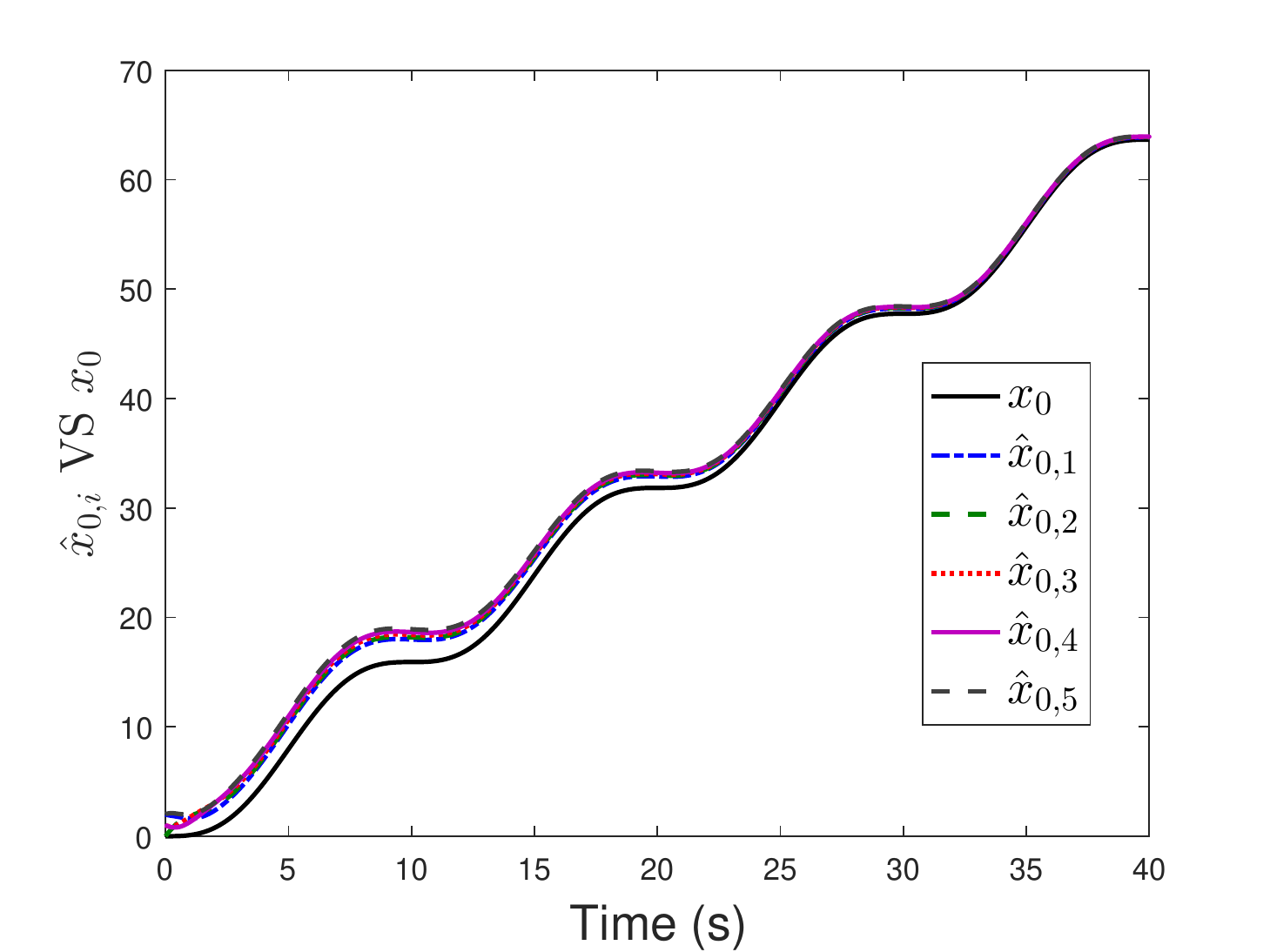}\label{x0-second}}\\
\subfloat[]{
\includegraphics[trim={6mm 1mm 12mm
 0mm},clip,width=0.38\linewidth]{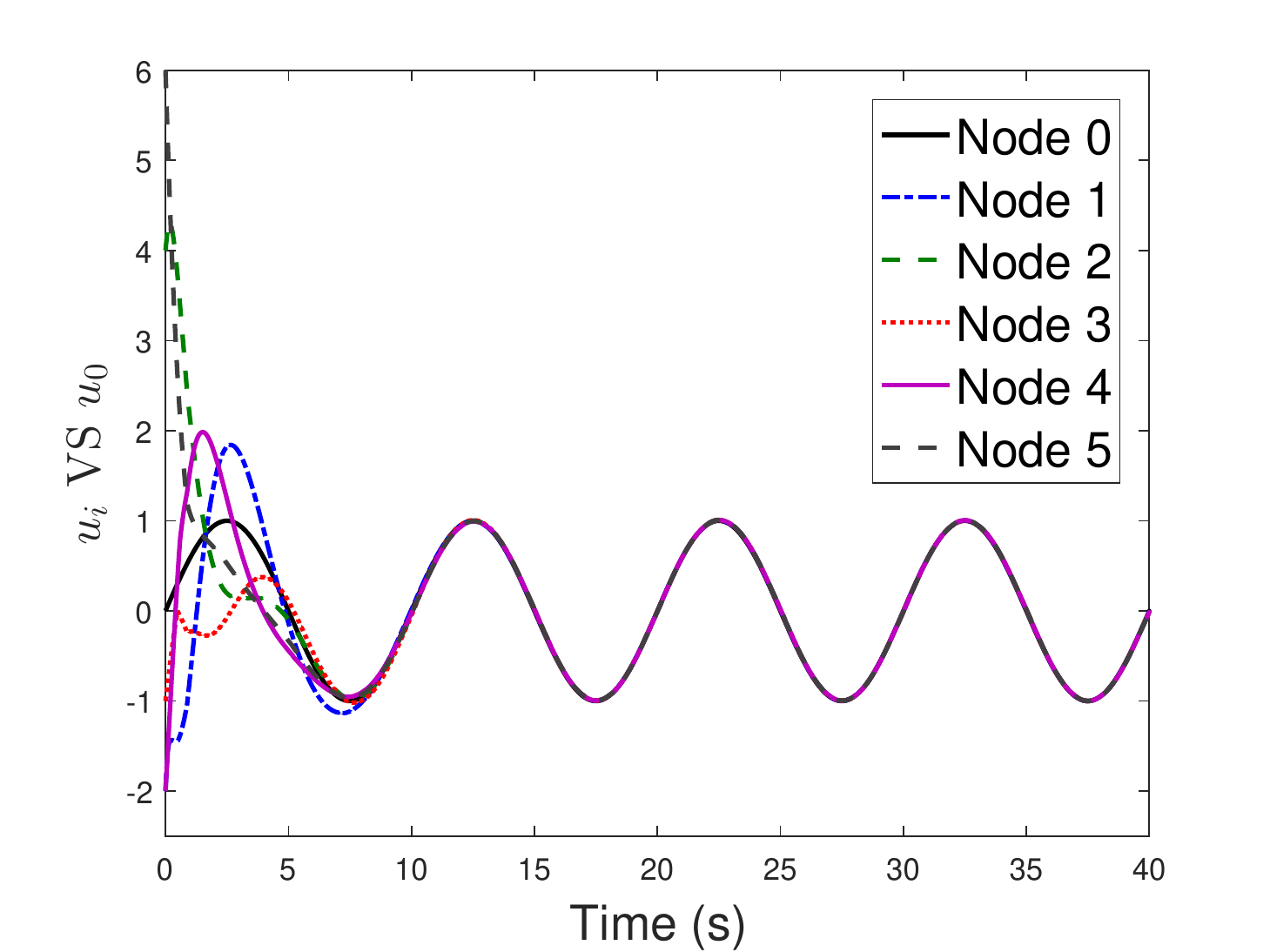}\label{ui-second}}
\vspace {-1mm}
\caption{Tracking control for a second-order a MAS: (a) position tracking; (b) velocity tracking; c) followers' estimation of the leader's velocity; (d)  follower's  estimation of their own velocities; (e) followers' estimation of the leader's input;
(f) followers' estimation of the leader's position; (g) followers' input  in comparison with the leader's.}\label{Second-order-MAS-profiles}
\end{figure*}
\section{NUMERICAL STUDY}\label{simulation}

In this section, we provide two illustrative examples to verify the effectiveness of the proposed distributed control algorithms. Consider an MAS consisting of one leader and five followers. The communication topology among them is shown in Figure~\ref{topology}. Node 0 is the leader, and nodes 1 to 5 are followers. The leader will only send information updates to follower 1, and the followers maintain undirectional communication with their neighbors. The corresponding Laplacian matrix  $L$ is given as follows:
\begin{align*}
L=\left[\begin{matrix}2 &-1& 0 &0 &-1\\
-1& 2& -1& 0& 0\\
0& -1& 2& -1& 0 \\
0& 0& -1& 2& -1\\
-1& 0& 0& -1& 2
\end{matrix}\right].
\end{align*}
Based on the communication topology, the diagonal matrix for the interconnection relationship between
the leader and the followers is
$
B=\mathrm{diag} (1, 0, 0, 0, 0)$. We choose $l=1$, $c=0.5$ and $\tau_i=1$ for $i=1,2,\ldots,N$.


We first consider the first-order tracking. The initial positions of the leader and followers are set to be $x(0)=\left[\begin{matrix} 0&3& 0 &-2& 1 &-1   \end{matrix}\right]^\top$. We assume that the leader's input profile is given as
$
u_0(t)=\sin(0.2\pi t)$.
The distributed tracking algorithm proposed in Section~\ref{first-order} is applied with the simulation results shown in Figure~\ref{First-order-MAS-profiles}. Figure~\ref{position-first} demonstrates the trajectories of the leader and followers as time goes by. It is seen that all the followers make an effort to track the leader from the beginning. After around 30 seconds, the followers catch up with the leader and keep an accurate tracking afterwards. Figure~\ref{u0-first} shows a comparison between the leader's actual input and the locally estimated input profiles by each follower. It demonstrates that the input estimation fast approaches the truth in the first three seconds and then maintains almost zero-error accuracy. Looking at the leader's position and the locally estimated profiles in Figure~\ref{x0-first}, one can see a good convergence of position estimation by the followers. The control input profiles of the followers are shown in Figure~\ref{ui-first}. For the leader and followers, their inputs gradually reach the same level after around 10 seconds, showing a synchronization in their maneuvers. 

We then consider the second-order tracking. The actual initial positions of the leader and followers are the same as in the previous case. Their initial velocities are $v(0)=\left[\begin{matrix} 0&1 &-2 &3& 0 &-1   \end{matrix}\right]^\top$. Figure~\ref{Second-order-MAS-profiles} summarizes the simulation results when the tracking algorithm in Section~\ref{second-order} is applied. Looking at the position trajectories of all followers and the leader in Figure~\ref{position-second}, one can see that all followers catch up with the leader after around 25 seconds and then well continue the tracking. Associated with
this position tracking, Figure~\ref{velocity-second} further illustrates the velocity tracking, which exhibits satisfactory convergence. The leader's velocity and the followers' estimation  are shown in Figure~\ref{leader-velocity-second}. It is seen that the velocity estimation by each follower converges to the truth at around the 12th second. Figure~\ref{follower-velocity-second} demonstrates that each follower begins to get accurate estimate of its own velocity at around the tenth second and then keeps an accurate estimation.
The time-based evolution of the leader's acceleration and its estimation by the followers is further shown in Figure~\ref{u0-second}. From this figure, the input observers of all the followers can capture the truth quickly in about three seconds, showing the effectiveness of estimation.
Figure~\ref{x0-second} illustrates the leader's position and the locally estimated profiles, between which there is a good agreement.
Finally, Figure~\ref{ui-second} shows the leader and followers' control input profiles, which gradually become the same. Through the above results and many others simulation runs, we consistently observe that the proposed input-observer-based tracking control algorithms can provide effective performance.

\section{Conclusion}\label{conclusion}

Leader-follower tracking represents an important task in diverse MAS mission contexts, which has been seeing a rapid rise of interest from researchers. In this paper, we proposed a novel input-observer-based  perspective into distributed tracking control design. Advancing the idea of observer-based tracking control in the literature, we highlighted that observers can be designed for a follower to directly estimate the leader's maneuver input and leverage the estimation to enhance tracking control. To this end, we developed distributed input observers along with some other observers and on such a basis, formulated a new tracking control framework. We conducted the study for both first- and second-order MASs, with a control approach developed for each case. We also pointed out that our approaches can help overcome a few limitations presented by some existing methods. We performed rigorous analysis to prove the convergence properties of the proposed approaches and further validate their effectiveness by numerical simulation.

\balance

\bibliographystyle{elsarticle-harv}        

\bibliography{MAS_Input_Observer_Design_ijc_R2_V2}

\end{document}